\documentstyle[aps,epsfig]{revtex}
\def\up{\uparrow}
\def\down{\downarrow}

\def\bfr{{\bf r}}
\def\bfA{{\bf A}}
\def\bfS{{\bf S}}
\begin{document}

\twocolumn[
 \hsize\textwidth\columnwidth\hsize
 \csname@twocolumnfalse\endcsname

\draft
\title{Spin-Dependent Josephson Current through Double Quantum Dots\\ and
 Measurement of Entangled Electron States}
\author{Mahn-Soo Choi, C. Bruder, and Daniel Loss}
\address{Department of Physics and Astronomy, University of Basel,
 Klingelbergstrasse 82, CH-4056 Basel, Switzerland}
\date{\today}
\maketitle

\begin{abstract}

We study a double quantum dot each dot of which is tunnel-coupled to
superconducting leads.  In the Coulomb blockade regime, a
spin-dependent Josephson coupling between two superconductors is
induced, as well as an antiferromagnetic Heisenberg exchange coupling
between the spins on the double dot which can be tuned by the
superconducting phase difference. We show that the correlated spin
states---singlet or triplets---on the double dot can be probed via the
Josephson current in a dc-SQUID setup.

\end{abstract}

%73.23.-b Mesoscopic systems
%73.23.Hk Coulomb blockade; single-electron tunneling
%74.50.+r Proximity effects, weak links, tunneling phenomena, and Josephson
%effects
]

In recent years, electronic transport through strongly interacting
mesoscopic systems has been the focus of many investigations
\cite{curacao}. In particular, a single quantum dot coupled via tunnel
junctions to two non-interacting leads has provided a prototype model
to study Coulomb blockade effects and resonant tunneling in such
systems. These studies have been extended to an Anderson impurity
\cite{Glazma89} or a quantum dot coupled to superconductors
\cite{Ralph95,Levy97,Rozhko99}. In a number of
experimental\cite{Ralph95} and theoretical \cite{Levy97} papers, the
spectroscopic properties of a quantum dot coupled to two
superconductors have been studied. Further, an effective dc Josephson
effect through strongly interacting regions between superconducting
leads has been analyzed \cite{Matveev93,Bauern94,Fazio96,Siewert96}.
More recently, on the other hand, research on the possibility to
control and detect the spin of electrons through their charges has
started. In particular in semiconducting nanostructures, it was found
that the direct coupling of two quantum dots by a tunnel junction can
be used to create entanglement between spins \cite{Lossxx98}, and that
such spin correlations can be observed

\begin{figure}\centering
\epsfig{file=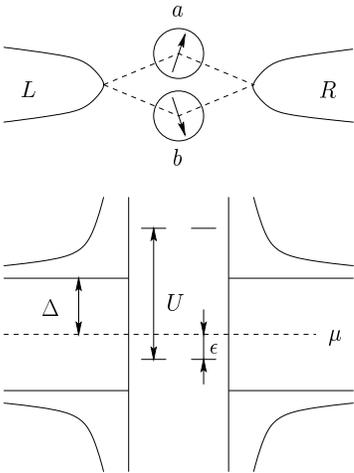,clip=,width=5.5cm}
\caption{Upper panel: sketch of the superconductor-double quantum
 dot-superconductor (S-DD-S) nanostructure. Lower panel: schematic
 representation of the quasiparticle energy spectrum in 
the superconductors and
 the single-electron levels of the two quantum dots.}
\label{sdds:fig1}
\end{figure}
\noindent
in charge transport experiments
%via two-particle interference effects
\cite{Lossxx99}.

Motivated by these studies we propose in the present work a new
scenario for inducing and detecting spin correlations, viz., coupling
a double quantum dot (DD) to superconducting leads by tunnel junctions
as shown in Fig.~\ref{sdds:fig1}. It turns out that this connection
via a superconductor induces a Heisenberg exchange coupling between
the two spins on the DD. Moreover, if the DD is arranged between two
superconductors (see Fig.~\ref{sdds:fig1}), we obtain a Josephson
junction (S-DD-S). The resulting Josephson current depends on the spin
state of the DD and can be used to {\it probe} the spin correlations
on the DD.

{\em Model\/}--- The double-dot (DD) system we propose is sketched in
Fig.~\ref{sdds:fig1}: Two quantum dots ($a$,$b$), each of which
contains one (excess) electron and is connected to two superconducting
leads ($L,R$) by tunnel junctions (indicated by dashed lines)
\cite{impurities}. There is no direct coupling between the two dots.
The Hamiltonian describing this system consists of three
parts, $H_S+H_D+H_T\equiv H_0+H_T$. The leads are assumed to be
conventional singlet superconductors that are described by the
BCS Hamiltonian
\begin{eqnarray} \label{sdds:1}
H_S
& = & \sum_{j=L,R}\int_{\Omega_j}{d\bfr\over \Omega_j}\;\Biggl\{
 \sum_{\sigma=\up,\down}
 \psi_\sigma^\dag(\bfr)h(\bfr)\psi_\sigma(\bfr)
 \\&&\mbox{}\qquad\qquad\nonumber
 + \Delta_j(\bfr)\psi_\up^\dag(\bfr)\psi_\down^\dag(\bfr)
 + h.c.
 \Biggr\} \;,
\end{eqnarray}
where $\Omega_j$ is the volume of lead $j$, 
$h(\bfr)
= (-i\hbar\nabla+\frac{e}{c}{\bf A})^2/2m
 - \mu$, and
$\Delta_j(\bfr)=\Delta_je^{-i\phi_j(\bfr)}$ is the pair potential. For
simplicity, we assume identical leads with same chemical
potential $\mu$, and $\Delta_L=\Delta_R=\Delta$.
The two quantum dots are modelled as two localized levels $\epsilon_a$
and $\epsilon_b$ with strong on-site Coulomb repulsion $U$, described by
the Hamiltonian
\begin{equation}
H_D
= \sum_{n=a,b}\left[
 - \epsilon\sum_{\sigma}d_{n\sigma}^\dag d_{n\sigma}
 + Ud_{n\up}^\dag d_{n\up}d_{n\down}^\dag d_{n\down}
\right] \;,
\end{equation}
where we put $\epsilon_a=\epsilon_b=-\epsilon$ ($\epsilon>0$) for
simplicity. $U$ is typically given by the charging energy of the dots,
and we have assumed that the level spacing of the dots is $\sim U$
(which is the case for small GaAs dots\cite{curacao}), so that we need to
retain only one energy level in $H_D$.
Finally, the DD is coupled {\em in parallel\/}
(see Fig.~\ref{sdds:fig1}) to the superconducting leads, described
by the tunneling Hamiltonian
\begin{equation} \label{sdds:5}
H_T
= \sum_{j,n,\sigma}\left[
 te^{-i\frac{\pi}{\Phi_0}
 \int_{\bfr_n}^{\bfr_{j,n}}d{\bf l}\cdot{\bf A}
 }\psi_\sigma^\dag(\bfr_{j,n}) d_{n\sigma}
 + h.c.
 \right] \;,
\end{equation}
where $\bfr_{j,n}$ is the point on the lead $j$ closest to the dot $n$.
Here,
$\Phi_0=hc/2e$ is the superconducting flux quantum.
Unless mentioned otherwise, it will be assumed that
$\bfr_{L,a}=\bfr_{L,b}=\bfr_L$ and
$\bfr_{R,a}=\bfr_{R,b}=\bfr_R$.

Since the low-energy states of the whole system are well separated by the
superconducting gap $\Delta$ as well as the strong Coulomb repulsion $U$
($\Delta,\epsilon\ll{}U-\epsilon$), it is sufficient to consider an
effective Hamiltonian on the reduced Hilbert space consisting of singly
occupied levels of the dots and the BCS ground states on the leads.
To lowest order in $H_T$, the effective Hamiltonian is
%using the projection method 
%\cite{Auerba94}
\begin{equation} \label{sdds:6}
H_{\it eff}
= P\, H_T\left[(E_0-H_0)^{-1}(1-P) H_T\right]^3\, P \; ,
%= P\, H_T\frac{1-P}{E_0-H_0}H_T\frac{1-P}{E_0-H_0}H_T
% \frac{1-P}{E_0-H_0}H_T\, P \;,
\end{equation}
where $P$ is the projection operator onto the subspace and
$E_0$ is the ground-state energy of the unperturbed Hamiltonian $H_0$.
(The 2nd order contribution leads to an irrelevant constant.)
The lowest-order expansion
(\ref{sdds:6})
is valid in the limit $\Gamma\ll\Delta,\epsilon$ where
$\Gamma=\pi{t^2}N(0)$ and $N(0)$ is
the normal-state density of states per spin of the

\begin{figure}\centering
\epsfig{file=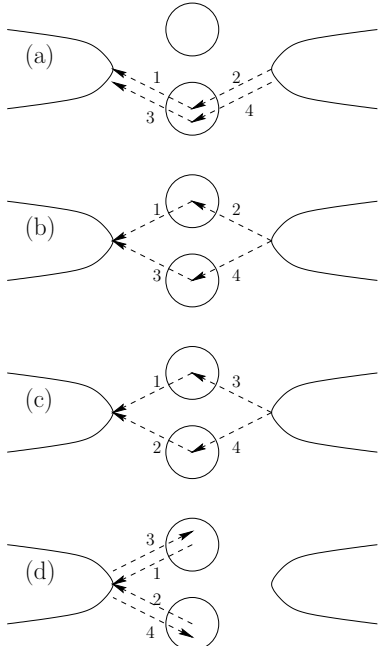,clip=,width=5cm}
\caption{Partial listing of virtual tunneling processes
 contributing to $H_{eff}$
(\ref{sdds:6}). The numbered
 arrows indicate the direction and the order of occurrence of the charge
 transfers.
 Processes of type (a) and (b) give a contribution proportional to $J_0$,
 whereas those of type (c) and (d) give contributions proportional to
 $J$. Other processes not listed here give negligible contributions in the
 energy regions of interest.}
\label{sdds:fig2}
\end{figure}

\noindent
leads at the Fermi energy. Thus, we assume that
$\Gamma\ll\Delta,\epsilon\ll U-\epsilon$, and temperatures which
are less than $\epsilon$ (but larger than the Kondo temperature).

{\em Effective Hamiltonian\/} ---
There are a number of virtual hopping processes
that contribute to the effective Hamiltonian (\ref{sdds:6}), see
Fig.~\ref{sdds:fig2} for a partial listing of them.
Collecting these various processes, one can get the effective Hamiltonian
in terms of the gauge-invariant phase differences $\phi$ and $\varphi$
between the superconducting leads and the spin operators $\bfS_a$ and
$\bfS_b$ of the dots (up to a constant and with $\hbar=1$)
\begin{eqnarray} \label{sdds:8}
H_{\it eff}
& = & J_0\cos(\pi{f}_{AB})\cos(\phi-\pi{f}_{AB}) \\
& + & \left[(2J_0+J)(1+\cos\varphi) + 2J_1(1+\cos\pi{f}_{AB})\right]
 \nonumber \\&&\mbox{}\times
 \left[\bfS_a\cdot\bfS_b-1/4\right] \;.
 \nonumber
\end{eqnarray}
Here $f_{AB}=\Phi_{AB}/\Phi_0$ and $\Phi_{AB}$ is the Aharonov-Bohm
(AB) flux threading through the closed loop indicated by the dashed lines
in Fig.~\ref{sdds:fig1}.
One should be careful to define {\em gauge-invariant\/} phase differences
$\phi$ and $\varphi$ in (\ref{sdds:8}). The phase difference $\phi$ is
defined as usual \cite{Tinkha96} by
\begin{equation}
\phi
= \phi_L(\bfr_L)-\phi_R(\bfr_R)
- \frac{2\pi}{\Phi_0}\int_{\bfr_R}^{\bfr_L}d\ell_a\cdot\bfA \;,
\end{equation}
where the integration from $\bfr_R$ to $\bfr_L$ runs via dot $a$
(see Fig.~\ref{sdds:fig1}).
The second phase difference, $\varphi$, is defined by
\begin{equation}
\varphi
= \phi_L(\bfr_L)-\phi_R(\bfr_R)
- \frac{\pi}{\Phi_0}\int_{\bfr_R}^{\bfr_L}(d\ell_a+d\ell_b)\cdot\bfA \;.
%- \frac{\pi}{\Phi_0}\int_{\bfr_R}^{\bfr_L}d\ell_b\cdot\bfA \;.
\end{equation}
The distinction between $\phi$ and $\varphi$, however, is not significant
unless one is interested in the effects of an AB flux through the 
closed loop in Fig.~\ref{sdds:fig1}
(see Ref.~\onlinecite{Lossxx99} for an example of such effects).
The coupling constants appearing in (\ref{sdds:8}) are defined by
\begin{eqnarray}
J
& = & \frac{2\Gamma^2}{\epsilon}\,
 \left[\frac{1}{\pi}\int\frac{dx}{f(x)g(x)}
 \right]^2 \nonumber \\
J_0
& = & \frac{\Gamma^2}{\Delta}
 \int\frac{dxdy}{\pi^2}\;
 \frac{1}{f(x)f(y)[f(x)+f(y)]g(x)g(y)} \\
J_1
& = & \frac{\Gamma^2}{\Delta}
 \int\frac{dxdy}{\pi^2}\;%\Biggl\{
 \frac{g(x)[f(x)+f(y)]-2\zeta g(y)}
 {g(x)^2g(y)[g(x)+g(y)][f(x)+f(y)]} \;,
 \nonumber
\end{eqnarray}
where $\zeta=\epsilon/\Delta$, $f(x)=\sqrt{1+x^2}$, and
$g(x)=\sqrt{1+x^2}+\zeta$.

Eq. (\ref{sdds:8}) is one of our main results. A remarkable feature of
it is that a Heisenberg exchange coupling between the spin on dot $a$
and on dot $b$ is induced by the superconductor. This coupling is
antiferromagnetic (all $J$'s are positive) and thus favors a singlet
ground state of spin $a$ and $b$. This in turn is a direct consequence
of the assumed singlet nature of the Cooper pairs in the
superconductor \cite{footnotetriplet}. As discussed below, an
immediate observable consequence of $H_{\it eff}$ is a {\em
spin-dependent\/} Josephson current from the left to right
superconducting lead (see Fig. \ref{sdds:fig1}) which 
probes the correlated spin state on the DD.

The various terms in (\ref{sdds:8}) have different magnitudes. In
particular, the processes leading to the $J_1$ term involve
quasiparticles only as can be seen from its AB-flux dependence which
has period $2 \Phi_0$. In the limits we will consider below, this
$J_1$ term is small and can be neglected.

In the limit $\zeta\gg{1}$, the main contributions come from processes of
the type depicted in Fig.~\ref{sdds:fig2} (a) and (b), making
$J_0\approx{}0.1(\Gamma^2/\zeta\epsilon)\log\zeta$ dominant over $J$
and $J_1$. Thus, (\ref{sdds:8}) can be reduced to
\begin{eqnarray} \label{sdds:10}
H_{\it eff}
& \approx & J_0\cos(\pi{f}_{AB})\cos(\phi-\pi{f}_{AB}) \\
& + & 2J_0(1 + \cos\varphi)\left[\bfS_a\cdot\bfS_b-\frac{1}{4}\right] \;,
 \nonumber
\end{eqnarray}
up to order $(\log\zeta)/\zeta$.
As can be seen in Fig.~\ref{sdds:fig2} (a), the first term in
(\ref{sdds:10}) has the same origin as that in the single-dot
case \cite{Glazma89}: Each dot separately constitutes an effective
Josephson junction with coupling energy $-J_0/2$ (i.e. $\pi$-junction)
between the two superconductors. The two resulting junctions form a dc
SQUID, leading to the total Josephson coupling in the first term of
(\ref{sdds:10}).
The Josephson coupling in the second term in (\ref{sdds:10}), corresponding
to processes of type Fig.~\ref{sdds:fig2} (b),
depends on the correlated spin states on the double dot: For the singlet
state,
it gives an ordinary Josephson junction with coupling $2J_0$
and competes with the first term, whereas it vanishes for the triplet
states.
Although the limit $\Delta\ll\epsilon\ll{}U-\epsilon$ is not easy to
achieve with present-day technology, such a regime is relevant, say,
for two atomic impurities embedded between the grains of
a granular superconductor.

More interesting and experimentally feasible is the case $\zeta\ll{1}$.
In this regime, the effective Hamiltonian (\ref{sdds:8}) is dominated by a
single term (up to terms of order $\zeta$),
\begin{equation} \label{sdds:12}
H_{\it eff}
\approx J(1+\cos\varphi)\;
 \left[{\bf S}_a\cdot{\bf S}_b-\frac{1}{4}\right] \;,
\end{equation}
with $J\approx{}2\Gamma^2/\epsilon$.
The processes of type Fig.~\ref{sdds:fig2} (b) and (c)
give rise to (\ref{sdds:12}).
Below we will propose an experimental setup based on (\ref{sdds:12}).

Before proceeding, we digress briefly on the dependence of $J$ on
the contact points. Unlike the processes of type Fig.~\ref{sdds:fig2} (a),
those of types Fig.~\ref{sdds:fig2} (b), (c), and (d) depend on
$\delta{r}_L=|\bfr_{L,a}-\bfr_{L,b}|$ and
$\delta{r}_R=|\bfr_{R,a}-\bfr_{R,b}|$, see the remark below
Eq.~(\ref{sdds:5}). For
the tunneling Hamiltonian (\ref{sdds:5}), one gets
(putting $\delta{r}=\delta{r}_L=\delta{r}_R$)
\begin{equation}
J(\delta{r})
= \frac{8t^4}{\epsilon}
 \left|\int_0^\infty\frac{d\omega}{2\pi}\;
 \frac{F^R(\delta{r},\omega)-F^A(\delta{r},\omega)}
 {\omega+\epsilon}
 \right|^2 \; ,
\end{equation}
where $F^{R/A}(\bfr,\omega)$ is the Fourier transform of
the Green's function in the superconductors,
$F^{R/A}(\bfr,t)=\mp i\Theta(\pm t)
\langle \{\psi_\up(\bfr,t),\psi_\down(0,0)\}\rangle$ 
\cite{footnotephases}.
 E.g., in the limit
$\varepsilon\ll\Delta\ll\mu$, we find
%one can calculate $J(\delta{r})$ explicitly
%to get
$J(\delta{r})
\approx J(0)e^{-2\delta{r}/\xi}
 \sin^2(k_F\delta{r})/(k_F\delta{r})^2$
up to order $1/k_F\xi$, with $k_F$ the Fermi wave vector
in the leads. Thus, to have $J(\delta{r})$ non-zero,
$\delta{r}$ should not exceed the
superconducting coherence length $\xi$.

{\em Probing spins with a dc-SQUID\/} ---
We now propose a possible experimental setup to probe the correlations
(entanglement) of the spins on the dots, based on the

\begin{figure}\centering
\epsfig{file=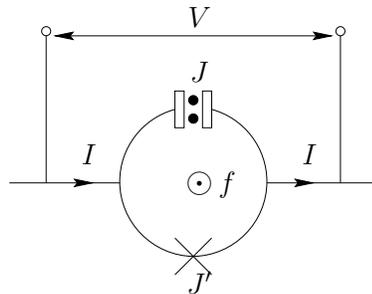,clip=,width=5cm}
\caption{dc-SQUID-like geometry consisting of 
the S-DD-S structure (filled dots at the
 top) connected in parallel with another ordinary Josephson junction
 (cross at the bottom).}
\label{sdds:fig3}
\end{figure}

\noindent
effective model
(\ref{sdds:12}). According to (\ref{sdds:12}) the S-DD-S structure
can be regarded as a {\em spin-dependent} Josephson
junction. Moreover, this structure can be connected with an ordinary
Josephson junction to form a dc-SQUID-like geometry, see
Fig.~\ref{sdds:fig3}. The Hamiltonian of the entire system is then
given by
\begin{eqnarray} \label{sdds:15}
H
& = & J[1+\cos(\theta-2\pi{f})]
 \left({\bf S}_a\cdot{\bf S}_b-\frac{1}{4}\right)
 \\&&\mbox{}\nonumber
 + \alpha J(1-\cos\theta) \;,
\end{eqnarray}
where $f=\Phi/\Phi_0$, $\Phi$ is the flux threading the SQUID loop,
$\theta$ is the gauge-invariant phase difference across the auxiliary
junction ($J'$), and $\alpha={}J'/J$ with $J'$ being the Josephson
coupling energy of the auxiliary junction\cite{footnotealpha}. One
immediate consequence of (\ref{sdds:15}) is that at zero temperature,
we can {\em effectively\/} turn on and off the spin exchange
interaction: For half-integer flux ($f=1/2$), singlet and triplet
states are degenerate at $\theta=0$. Even at finite temperatures,
where $\theta$ is subject to thermal fluctuations, singlet and triplet
states are almost degenerate around $\theta=0$. On the other hand,
for integer flux ($f=0$), the energy of the singlet is lower by $J$
than that of the triplets.

\begin{figure}\centering
\epsfig{file=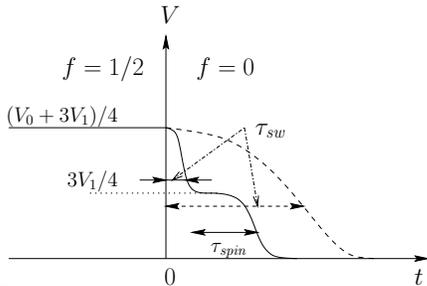,clip=,width=6cm}
\caption{Schematic representation of dc voltage $V$ vs.
time when probing the spin correlations of the DD. The flux through 
the SQUID loop is switched from $f=1/2$ to $f=0$ at $t=0$.
Solid line: $\tau_{\it sw} < \tau_{\it spin}$. Dashed line:
$\tau_{\it sw} > \tau_{\it spin}$.
}
\label{sdds:fig4}
\end{figure}

This observation allows us to probe directly the spin state on the
double dot via a Josephson current across the dc-SQUID-like structure in
Fig.~\ref{sdds:fig3}. 
The supercurrent through the SQUID-ring is defined as
$I_S=(2\pi c/\Phi_0)\partial \langle H\rangle /\partial\theta$,
where the brackets refer to a spin expectation value on the DD.
Thus, depending on the spin state on the DD we find
\begin{equation} \label{sdds:16}
I_S/I_J
= \left\{\begin{array}{ll}
 \sin(\theta-2\pi{f}) + \alpha\sin\theta
 &\mbox{(singlet)} \\
 \alpha\sin\theta
 &\mbox{(triplets)}
 \end{array}\right. \;,
\end{equation}
where $I_J = 2eJ/\hbar$. When the system is biased by a dc current
$I$ larger than the spin- and flux-dependent critical current, given
by $\max_{\theta} \{|I_S|\}$, a finite voltage $V$ appears. Then one
possible experimental procedure might be as follows (see
Fig.~\ref{sdds:fig4}). Apply a dc bias current such that $\alpha I_J
< I < (\alpha+1) I_J$. Here, $\alpha{I_J}$ is the critical current of
the triplet states, and $(\alpha+1)I_J$ the critical current of the
singlet state at $f=0$, see (\ref{sdds:16}). Initially prepare the
system in an equal mixture of singlet and triplet states by tuning the
flux around $f=1/2$. (With electron $g$-factors $g\sim0.5\mbox{--}20$
the Zeeman splitting on the dots is usually small compared with $k_BT$
and can thus be ignored.) The dc voltage measured in this mixture
will be given by $(V_0+3V_1)/4$, where $V_0 (V_1) \sim 2\Delta/e$ 
is the (current-dependent) voltage drop associated with the
singlet (triplet) states. At a later time $t=0$, the
flux is switched off (i.e. $f=0$), with $I$ being kept fixed. The
ensuing time evolution of the system is characterized by three time
scales: the time $\tau_{\it coh} \sim \max\{1/\Delta,1/\Gamma\}\sim
1/\Gamma$ it takes to establish coherence in the S-DD-S junction, the
spin relaxation time $\tau_{\it spin}$ on the dot, and the switching
time $\tau_{\it sw}$ to reach $f=0$. We will assume $\tau_{\it coh}
\ll \tau_{\it spin}, \tau_{\it sw}$, which is not unrealistic in view
of measured spin decoherence times in GaAs exceeding $100$ ns
\cite{Awschalom}. If $\tau_{\it sw} < \tau_{\it spin}$, the voltage
is given by $3V_1/4$ for times less than $\tau_{\it spin}$, i.e. the
singlet no longer contributes to the voltage. For $t>\tau_{\it spin}$
the spins have relaxed to their ground (singlet) state, and the
voltage vanishes. One therefore expects steps in the voltage versus
time (solid curve in Fig.~\ref{sdds:fig4}). If $\tau_{\it
spin}<\tau_{\it sw} $, a broad transition region of the voltage from
the initial value to $0$ will occur (dashed line in
Fig.~\ref{sdds:fig4}) \cite{footnoterfSQUID}.

To our knowledge, there are no experimental reports on quantum dots
coupled to superconductors. However, hybrid systems
consisting of superconductors (e.g., Al or Nb) and 2DES (InAs and
GaAs) have been investigated by a number of groups \cite{vanWees}.
Taking the parameters of those materials, a rough estimate leads to a
coupling energy $J$ in (\ref{sdds:12}) or (\ref{sdds:15}) of about
$J\sim\mbox{0.05--0.5K}$. This corresponds to a critical current scale of
$I_J\sim\mbox{5--50}\mbox{nA}$.

In conclusion, we have investigated double quantum dots each dot of
which is coupled to two superconductors. We have found that in the
Coulomb blockade regime the Josephson current from one superconducting
lead to the other is different for singlet or triplet states on the
double dot. This leads to the possibility to probe the spin states of
the dot electrons by measuring a Josephson current.

{\em Acknowledgment\/} ---
We would like to thank G. Burkard, C. Strunk, and E. Sukhorukov for
discussions and the Swiss NSF for support.


\begin{thebibliography}{1}
\bibitem{impurities} 
One could also consider atomic impurities
embedded between the grains of a granular superconductor.
\bibitem{footnotephases}
We note that the phase difference $\varphi$ in
(\ref{sdds:8}) should now be defined with respect to the phase
$\phi_j(\bfr_{j,a},\bfr_{j,b})$ of the function
$F_j^R(\bfr_{j,a},\bfr_{j,b})-F_j^A(\bfr_{j,a},\bfr_{j,b})$
on the lead $j$, see the definition below (\ref{sdds:8}).
\bibitem{footnotealpha}
Without restriction we can assume $\alpha>1$, since
$J'$ could be adjusted accordingly
by replacing the $J'$-junction by
another dc SQUID and flux through it.
\bibitem{footnoterfSQUID}
Another experimental setup would be to use an rf-SQUID
geometry, i.e., to embed the S-DD-S structure into a superconducting
ring \cite{Tinkha96}. However, to operate such a device, ac fields are
necessary, and the sensitivity is not as good as for
the dc-SQUID geometry.
\end{thebibliography}
\end{document}